# Secured Cryptographic Key Generation From Multimodal Biometrics: Feature Level Fusion of Fingerprint and Iris

A.Jagadeesan
Research scholar/Senior Lecturer/EIE
Bannari Amman Institute of Technology
Sathyamangalam-638 401, Tamil Nadu, India
.

Dr. K.Duraiswamy
Dean/Academic
K.S.Rangasamy College of Technology,
Tiruchengode – 637 209, Tamil Nadu, India
.

*Abstract*— **Human users have a tough time remembering long cryptographic keys. Hence, researchers, for so long, have been examining ways to utilize biometric features of the user instead of a memorable password or passphrase, in an effort to generate strong and repeatable cryptographic keys. Our objective is to incorporate the volatility of the user's biometric features into the generated key, so as to make the key unguessable to an attacker lacking significant knowledge of the user's biometrics. We go one step further trying to incorporate multiple biometric modalities into cryptographic key generation so as to provide better security. In this article, we propose an efficient approach based on multimodal biometrics (Iris and fingerprint) for generation of secure cryptographic key. The proposed approach is composed of three modules namely, 1) Feature extraction, 2) Multimodal biometric template generation and 3) Cryptographic key generation. Initially, the features, minutiae points and texture properties are extracted from the fingerprint and iris images respectively. Subsequently, the extracted features are fused together at the feature level to construct the multi-biometric template. Finally, a 256-bit secure cryptographic key is generated from the multi-biometric template. For experimentation, we have employed the fingerprint images obtained from publicly available sources and the iris images from CASIA Iris Database. The experimental results demonstrate the effectiveness of the proposed approach.**

I. INTRODUCTION

The necessity for reliable user authentication techniques has risen amidst of heightened issues about security and rapid progress in networking, communication and mobility [1]. The generally utilized authentication systems that regulate the entry to computer systems or secured locations are password, but it can be cracked or stolen. For that reason, biometrics has turned out to be a practicable option to traditional identification methods in several application areas [23]. Biometrics, expressed as the science of identifying an individual on the basis of her physiological or behavioral traits, seems to achieve acceptance as a rightful method for obtaining an individual's identity [1]. Biometric technologies have established their importance in a variety of security, access control and monitoring applications. The technologies are still novel and momentarily evolving [2]. Biometric systems possess numerous advantages over traditional authentication methods, that is, 1) biometric information cannot be obtained by direct covert observation, 2) It is difficult to share and reproduce, 3) It improves user easiness by lessening the necessity to memorize long and random passwords, 4) It safeguards against repudiation by the user. Besides, biometrics imparts the same security level to all users unlike passwords and is tolerant to brute force attacks [3]. A number of biometric characteristics are being employed today, which comprises fingerprint, DNA, iris pattern, retina, ear, thermogram, face, gait, hand geometry, palm-vein pattern, smell, keystroke dynamics, signature, and voice [16, 17].

Biometric systems that generally employ a single attribute for recognition (that is., unimodal biometric systems) are influenced by some practical issues like noisy sensor data, non-universality and/or lack of distinctiveness of the biometric trait, unacceptable error rates, and spoof attacks [4]. A probable improvement, multimodal biometric systems prevail over some of these issues by strengthening the proof acquired from several sources [5] [6]. Multimodal biometric system employs two or more individual modalities, namely, gait, face, Iris and fingerprint, to enhance the recognition accuracy of conventional unimodal methods. With the use of multiple biometric modalities, it is shown that to decrease error rates, by offering extra valuable information to the classifier. Diverse characteristics can be employed by a single system or separate systems that can function on its own and their decisions may be merged together [7]. The multimodal-based authentication can aid the system in improving the security and effectiveness in comparison to unimodal biometric authentication, and it might become challenging for an adversary to spoof the system owing to two individual biometrics traits [15].

In recent times, multimodal biometrics fusion techniques have invited considerable attention as the supplementary information between different modalities could enhance the recognition performance. Majority of the works have focused the attention in this area [8-10]. In most cases, they can be categorized into three groups: fusion at the feature level, fusion at the match level and fusion at the decision level [6] [11]. Fusion at the feature level includes the incorporation of feature sets relating to multiple modalities. The feature set holds richer information about the raw biometric data than the match score or the final decision and because of this, integration at this level is bound to offer good recognition results. But, fusion at this level is hard to accomplish in real time due to the following grounds: (i) the feature sets of multiple modalities may be unsuited (namely, minutiae set of fingerprints and eigen-coefficients of face); (ii) the association between the feature spaces of diverse biometric systems may be unknown; and (iii) concatenating two feature vectors may lead to a feature vector with very high dimensionality resulting to the `curse of dimensionality' problem [12].

One current development, biometric cryptosystems [13] join cryptography and biometrics to take advantage from the strong points of both fields. In such systems, while cryptography endows with high and modifiable security levels, biometrics provides non-repudiation and removes the requirement to memorize passwords or to carry tokens [14]. Lately, the improved performance of cryptographic key generated from biometrics in accordance to security has acquired massive reputation amongst the researchers and experimenters [18] and recently, researchers have made an effort towards combing biometrics with cryptography so as to enhance the security, by removing the requirement for key storage using passwords [19-22]. Although it is highly impractical to break cryptographic keys generated from biometrics, the attackers have a good possibility of stealing by cryptographic attacks. One effectual solution with additional security will be the integration of multimodal biometrics into cryptographic key generation; in order to attain incredible security against cryptographic attacks.

At this juncture, we introduce an efficient approach for the secure cryptographic key generation on the basis of multiple modalities like, Iris and fingerprint. At first, the fingerprint features (minutiae points) are obtained from the fingerprint image using segmentation, Orientation field estimation and morphological operators. Likewise, the texture features are acquired from the iris image by segmentation, estimation of iris boundary and Normalization. Minutiae points and iris texture, the two extracted features are then fused at feature level to build the multimodal biometric template. Fusion at the feature level is achieved by means of the processes that is, concatenation, shuffling and merging. Lastly, multi-biometric template acquired is employed to generate the secure 256-bit cryptographic key that is competent of enabling better user authentication and security.

The rest of the paper is organized as follows. A short review of the researches connected to the proposed approach is given in Section II. The proposed approach for generation of multimodal-based cryptographic key is demonstrated in Section III. The results acquired on experimentation of the proposed approach are given in Section IV. To conclude, the conclusions are summed up in Section V.

II. REVIEW OF RELATED RESEARCHES

Literature embrace ample researches for generating cryptographic keys from biometric modalities and multimodal biometrics based user authentication. Amid all these researches, approaches for cryptographic key generation from biometric features and authenticating users by combining multiple biometric modalities, comprise a grand consideration in the recent development. A brief review of some recent researches is presented here.

A realistic and safe approach to incorporate the iris biometric into cryptographic applications has been presented by Feng Hao *et al.* [31].This approach employed a recurring binary string, called as a biometric key, that was created from a subject's iris image with the help of auxiliary error-correction data, which does not disclose the key and can be stored in a tamper-resistant token, like a smart card. The reproduction of the key revolves on two aspects: the iris biometric and the token. The assessment was done by using iris samples from 70 different eyes, with 10 samples from each eye. This resulted with the genuine iris codes with a 99.5 percent achievement rate, which upshot with 140 bits of biometric key which is sufficient for a 128-bit AES.A technique presented by B. Chen and V. Chandran [21], coalesced entropy based feature extraction process with Reed-Solomon error correcting codes which generate deterministic bit-sequences from the output of an iterative one-way transform. The technique was assessed using 3D face data and was proved to generate keys of suitable length for 128-bit Advanced Encryption Standard (AES).

A biometric-key generation scheme based on a randomized biometric helper has been presented by Beng.A *et al.* [42]. The technique engrosses a randomized feature discretization process and a code redundancy construction. The former method controls the intra-class variations of biometric data to the nominal level and the latter reduced the errors even more. The randomized biometric feature was proved as a simple technique, when the key was conciliated The projected technique was assessed in the context of face data based on a subset of the Facial Recognition Technology (FERET) database. Sanaul Hoque *et al.* [43] have presented the direct generation of the biometric keys from live biometrics, under certain conditions, by partitioning feature space into subspaces and partitioning these into cells, where each cell subspace contributes to the overall key generated. They assessed the presented technique on real biometric data, instead of both genuine samples and attempted imitations. Experimental results have proved the reliability in possible practical scenarios for this technique.

A cryptographic key generation from biometric data, based on lattice mapping based fuzzy commitment method was proposed by Gang Zheng *et al.* [44].This was proposed with the aim to, secure the biometric data even when the stored information in the system was open to an attacker ,with the generation of high entropy keys and also concealed the original biometric data. Simulated results have proved that its authentication accuracy was on par to the k-nearest neighbor classification. Tianhao Zhang *et al.* [45] have presented a Geometry Preserving Projections (GPP) method for subspace selection.It is capable of discriminating different classes and conserving the intra-modal geometry of samples within an identical class. With GPP, they projected all raw biometric data from different identities and modalities onto a unified subspace, on which classification can be executed. Also, the training stage was performed after having a unified transformation matrix to project different modalities. Experimental results have proved the effectiveness of the presented GPP for individual recognition tasks.

Donald E. Maurer and John P. Baker *et al have presented a* fusion architecture based on Bayesian belief networks [46]. The technique fully exploited the graphical structure of Bayes nets to define and explicitly model statistical dependencies between relevant variables: per sample measurements like, match scores and corresponding quality estimates and global decision variables. These statistical dependencies are in the form of conditional distributions which are modeled as Gaussian, gamma, log-normal or beta.. Each model is determined by its mean and variance, thus considerably minimizing training data needs. Furthermore, they retrieved the information from lower quality measurements by conditioning decision variables on quality as well as match score instead of rejecting them out of hand. Another characteristic of the technique was a global quality measure intended to be used as a confidence estimate supporting decision making. Introductory studies using the architecture to fuse fingerprints and voice were accounted.

Muhammad Khurram Khana and Jiashu Zhanga presented an efficient multimodal face and fingerprint biometrics authentication system on space-limited tokens, e.g. smart cards, driver license, and RFID cards [47]. Fingerprint templates were encrypted and encoded/embedded within the face images in a secure manner, so that the feature does not get changed drastically during encoding and decoding. This method of biometrics authentication on space-limited tokens, was proven to be a proficient and a cheap alternative without downgrading the overall decoding and matching performance of the biometrics system. A class-dependence feature analysis technique based on Correlation Filter Bank (CFB) technique for efficient multimodal biometrics fusion at the feature level is presented by Yan Yan and Yu-Jin Zhang [48]. In CFB, by optimizing the overall original correlation outputs the unconstrained correlation filter trained for a specific modality. So, the variation between modalities has been considered and the useful information in various modalities is completely utilized. Previous experimental outcome on the fusion of face and palmprint biometrics proved the advantage of the technique.

An authentication method presented by M.Nageshkumar *et al.* [24], focuses *on multimodal* biometric system identification using two features i.e. face and palmprint. The technique was produced for application where the training data includes a face and palmprint. Mixing the palmprint and face features has enhanced the robustness of the person authentication. The final assessment was done by fusion at matching score level architecture where features vectors were formed independently for query measures and are then evaluated to the enrolment template, which were saved during database preparation. Multimodal biometric system was expanded through fusion of face and palmprint recognition.

### III. PROPOSED APPROACH FOR CRYPTOGRAPHIC KEY GENERATION FROM MULTIMODAL BIOMETRICS

Multimodal biometric authentication has been more reliable and capable than knowledge-based (e.g. Password) and token-based (e.g. Key) techniques and has recently emerged as an attractive research area [24]. Several researchers [45-48] have successfully made use of multiple biometric traits for achieving user authentication. Multimodal biometrics was aimed at meeting the stringent performance requirements set by security-conscious customers. Some good advantages of multimodal biometrics are 1) improved accuracy 2) secondary means of enrollment and verification or identification in case sufficient data is not extracted from a given biometric sample and 3) ability to detect attempts to spoof biometric systems through non-live data sources such as fake fingers. Two important parameters that determine the effectiveness of the multimodal biometrics are choice of the biometric traits to be combined and the application area. The different biometric traits include fingerprint, face, iris, voice, hand geometry, palmprint and more. In the proposed approach, we integrate fingerprint and iris features for cryptographic key generation. The use of multimodal biometrics for key generation provides better security, as it is made difficult for an intruder to spool multiple biometric traits simultaneously. Moreover, the incorporation of biometrics into cryptography shuns the need to remember or carry long passwords or keys. The steps involved in the proposed multimodal-based approach for cryptographic key generation are,

1) Feature extraction from fingerprint.
2) Feature extraction from iris.
3) Fusion of fingerprint and iris features.
4) Generation of cryptographic key from fused features.

*A. Minutiae Points Extraction from Fingerprints*

This sub-section describes the process of extracting the minutiae points from the fingerprint image. We chose fingerprint biometrics chiefly because of its two significant characteristics: uniqueness and permanence (ability to remain unchanged over the lifetime). A fingerprint can be described as a pattern of ridges and valleys found on the surface of a fingertip. The ridges of the finger form the so-called minutiae points: ridge endings (terminals of ridge lines) and ridge bifurcations (fork-like structures) [26]. These minutiae points serve as an important means of fingerprint recognition. The steps involved in the proposed approach for minutiae extraction are as follows,

*1) Preprocessing:* The fingerprint image is first preprocessed by using the following methods,
- Histogram Equalization
- Wiener Filtering

**Histogram Equalization:** Histogram equalization (HE) is a very common technique for enhancing the contrast of an image. Here, the basic idea is to map the gray levels based on the probability distribution of the input gray levels. HE flattens and stretches the dynamic range of the image's histogram resulting in overall contrast improvement of the image [32]. HE transforms the intensity values of the image as given by the equation,

$$s_k = T(r_k) = \sum_{j=1}^{k} P_r(r_j) = \sum_{j=1}^{k} \frac{n_j}{n}$$

Where $s_k$ is the intensity value in the processed image corresponding to intensity $r_k$ in the input image, and $p_r(r_j) = 1,2,3....L$ is the input fingerprint image intensity level [33].

*Wiener filtering:* Wiener filtering improves the legibility of the fingerprint without altering its ridge structures [34]. The filter is based on local statistics estimated from a local neighborhood $\eta$ of size $3 \times 3$ of each pixel, and is given by the following equation:

$$w(n_1, n_2) = \mu + \frac{\sigma^2 - v^2}{\sigma^2}(I(n_1, n_2) - \mu)$$

where $v^2$ is the noise variance, $\mu$ and $\sigma^2$ are local mean and variance and $I$ represents the gray level intensity in $n_1, n_2 \in \eta$ [35].

*2) Segmentation:* The fingerprint image obtained after preprocessing is of high contrast and enhanced visibility. The next step is to segment the preprocessed fingerprint image. First, the fingerprint image is divided into non-overlapping blocks of size 16x16. Subsequently, the gradient of each block is calculated. The standard deviation of gradients in X and Y direction are then computed and summed. If the resultant value is greater than the threshold value the block is filled with ones, else the block is filled with zeros.

*3) Orientation Field Estimation:* A fingerprint orientation field is defined as the local orientation of the ridge-valley structures [27]. To obtain reliable ridge orientations, the most common approach is to go through the gradients of gray intensity. In the gradient-based methods, gradient vectors $[g_x, g_y]^T$ are first calculated by taking the partial derivatives of each pixel intensity in Cartesian coordinates. Traditional gradient-based methods divide the input fingerprint image into equal-sized blocks of $N \times N$ pixels, and average over each block independently [25] [26]. The direction of orientation field in a block is given by,

$$\theta_B = \frac{1}{2} a\tan\left(\frac{\sum_{i=1}^{N}\sum_{j=1}^{N} 2g_x(i,j)g_y(i,j)}{\sum_{i=1}^{N}\sum_{j=1}^{N} g_x^2(i,j) - g_y^2(i,j)}\right) + \frac{\pi}{2}$$

Note that function $a\tan(\cdot)$ gives an angle value ranges in $(-\pi, \pi)$ which corresponds to the squared gradients, while $\theta_B$ is the desired orientation angle within $[0, \pi]$.

*4) Image Enhancement:* It would be desirable to enhance the fingerprint image further prior to minutiae extraction. The fingerprint image enhancement is achieved by using,
- Gaussian Low-Pass Filter
- Gabor Filter

*Gaussian Low-Pass Filter:* The Gaussian low-pass filter is used as to blur an image. The Gaussian filter generates a `weighted average' of each pixel's neighborhood, with, the average weighted more towards the value of the central pixels. Because of this, gentler smoothing and edge preserving can be achieved. The Gaussian filter uses the following 2-D distribution as a point-spread function, and is achieved by the convolution [28].

$$G(x, y) = \left(\frac{1}{2\pi\sigma}\right)^2 \exp\left\{\frac{(x^2 + y^2)}{2\sigma^2}\right\}$$

Where, $\sigma$ is the standard deviation of the distribution.

*Gabor Filter:* Mostly used contextual filter [29] for fingerprint image enhancement is Gabor filter proposed by Hong, Wan, and Jain [30]. Gabor filters have both frequency-selective and orientation-selective properties and they also have optimal joint resolution in both spatial and frequency domains. The following equation shows the 2-Dimensional (2-D) Gabor filter form [29],

$$G(x, y, \theta, f_0) = \exp\left\{-\frac{1}{2}\left(\frac{x_\theta^2}{\sigma_x^2} + \frac{y_\theta^2}{\sigma_y^2}\right)\right\}\cos(2\pi f_0 x_\theta),$$

$$\begin{bmatrix} x_\theta \\ y_\theta \end{bmatrix} = \begin{bmatrix} \sin\theta & \cos\theta \\ -\cos\theta & \sin\theta \end{bmatrix}\begin{bmatrix} x \\ y \end{bmatrix}$$

where $\theta$ is the orientation of the filter, $f_0$ is the ridge frequency, $[x_\theta, y_\theta]$ are the coordinates of $[x, y]$ after a clockwise rotation of the Cartesian axes by an angle of $(90° - \theta)$, and $\sigma_x$ and $\sigma_y$ are the standard deviations of the Gaussian envelope along the $x$-and $y$-axes, respectively.

*5) Minutiae extraction:* The process of minutiae point extraction is carried out in the enhanced fingerprint image. The steps involved in the extraction process are,
- Binarization
- Morphological Operators

*Binarization:* Binarization is the process of converting a grey level image into a binary image. It improves the contrast between the ridges and valleys in a fingerprint image, and thereby facilitates the extraction of minutiae. The grey level value of each pixel in the enhanced image is examined in the binarization process. If the grey value is greater than the global threshold, then the pixel value is set to a binary value one; or else, it is set to zero. The output of binarization process is a binary image containing two levels of information, the foreground ridges and the background valleys. The minutiae extraction algorithms are good operating on binary images where there are only two levels of interest: the black pixels that denote ridges, and the white pixels that denote valleys.

*Morphological Operations:* Following the binarization process, morphological operators are applied to the binarized fingerprint image. The objective of the morphological operations is to eliminate obstacles and noise from the image. Furthermore, the unnecessary spurs, bridges and line breaks are removed by these operators. The process of removal of redundant pixels till the ridges become one pixel wide is facilitated by ridge thinning. The Ridge thinning algorithm utilized for Minutiae points' extraction in the proposed approach has been employed by the authors of [36]. The image is first divided into two dissimilar subfields that resemble a checkerboard pattern. In the first sub iteration, the pixel p from the initial subfield is erased only when all three conditions, G1, G2, and G3 are satisfied. While, in the second sub iteration, the pixel p from the foremost subfield is erased when all three conditions, G1, G2, and G3' are satisfied.

**Condition G1:**
$X_H(P) = 1$
Where
$X_H(P) = \sum_{i=1}^{4} b_i$
$b_i = \begin{cases} 1 \text{ if } x_{2i-1} = 0 \text{ and } (x_{2i} = 1 \text{ or } x_{2i+1} = 1) \\ 0 \text{ otherwise} \end{cases}$

$x_1, x_2, ..., x_8$ are the values of the eight neighbors of $p$, starting with the east neighbor and numbered in counter-clockwise order.

**Condition G2:**
$2 \leq \min\{n_1(p), n_2(p)\} \leq 3$
where
$n_1(p) = \sum_{k=1}^{4} x_{2k-1} \vee x_{2k}$
$n_2(p) = \sum_{k=1}^{4} x_{2k} \vee x_{2k+1}$

**Condition G3:**
$(x_2 \vee x_3 \vee \bar{x}_8) \wedge x_1 = 0$

**Condition G3':**
$(x_6 \vee x_7 \vee \bar{x}) \wedge x_5 = 0$

The resultant fingerprint image produced by the morphological thinning algorithm composes of ridges each one pixel wide. This improves the visibility of the ridges and enables effective and effortless of minutiae points.

*B. Feature Extraction from Iris*

The process of extracting features from the iris image is discussed in this sub-section. Iris recognition has been recognized as an effective means for providing user authentication. One important characteristic of the iris is that, it is so unique that no two irises are alike, even among identical twins, in the entire human population [37]. The human iris, an annular part between the pupil (generally, appearing black in an image) and the white sclera has an extraordinary structure and offers a plenty of interlacing minute characteristics such as freckles, coronas, stripes and more. These visible characteristics, which are generally called the texture of the iris, are unique to each subject [38]. The steps involved in the feature extraction process of the iris image are given below.

*1) Segmentation:* Iris segmentation is an essential module in iris recognition because it defines the effective image region used for subsequent processing such as feature extraction. Generally, the process of iris segmentation is composed of two steps 1) Estimation of iris boundary and 2) Noise removal.

*Estimation of iris boundary:* For boundary estimation, the iris image is first fed to the canny algorithm which generates the edge map of the iris image. The detected edge map is then used to locate the exact boundary of pupil and iris using Hough transform.

- *Canny edge detection*
The Canny edge detection operator was developed by John F. Canny in 1986. It uses a multi-stage algorithm to detect a wide range of edges in images. Canny edge detection starts with linear filtering to compute the gradient of the image intensity distribution function and ends with thinning and thresholding to obtain a binary map of edges. One significant feature of the Canny operator is its optimality in handling noisy images as the method bridges the gap between strong and weak edges of the image by connecting the weak edges in the output only if they are connected to strong edges. Therefore, the edges will probably be the actual ones. Hence compared to other edge detection methods, the canny operator is less fooled by spurious noise [39].

- *Hough Transform*
The classical Hough transform was concerned with the identification of lines in the image, but later, the Hough transform has been extended to identify positions of arbitrary shapes, most commonly circles or ellipses. From the edge map obtained, votes are cast in Hough space for the parameters of circles passing through each edge point. These parameters are the centre coordinates *x* and *y*, and the radius *r*, which are able to define any circle according to the equation,

$$x^2 + y^2 = r^2$$

A maximum point in the Hough space will correspond to the radius and centre coordinates of the circle best defined by the edge points.

*Isolation of Eyelids and Eyelashes:* In general, the eyelids and eyelashes occlude the upper and lower parts of the iris region. In addition, specular reflections can occur within the iris region corrupting the iris pattern. The removal of such noises is also essential for obtaining reliable iris information.

• Eyelids are isolated by fitting a line to the upper and lower eyelid using the linear Hough transform. A second horizontal line is then drawn, which intersects with the first line at the iris edge that is closest to the pupil; the second horizontal line allows maximum isolation of eyelid region.

• The eyelashes are quite dark compared with the surrounding eyelid region. Therefore, thresholding is used to isolate eyelashes.

*2) Iris Normalization:* Once the iris image is efficiently localized, then the next step is to transform it into the rectangular sized fixed image. The transformation process is carried out using the Daugman's Rubber Sheet Model.

***Daugman's Rubber Sheet Model:*** Normalization process involves unwrapping the iris and converting it into its polar equivalent. It is done using Daugman's Rubber sheet model [40] and is shown in figure.

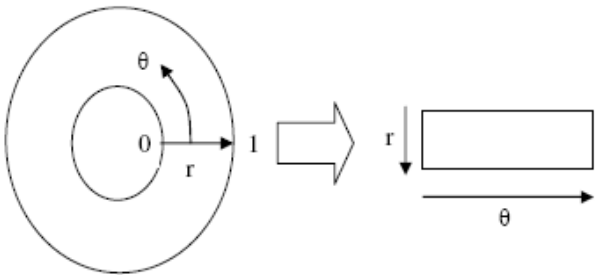

Figure 1. Daugman's Rubber Sheet Model

For every pixel in the iris, an equivalent position is found out on polar axes. The process comprises of two resolutions: Radial resolution, which is the number of data points in the radial direction and Angular resolution, which is the number of radial lines generated around iris region. Using the following equation, the iris region is transformed to a 2D array with horizontal dimensions of angular resolution and vertical dimension of radial resolution.

$$I[x(r,\theta), y(r,\theta)] \rightarrow I(r,\theta)$$

where, $I(x, y)$ is the iris region, $(x, y)$ and $(r, \theta)$ are the Cartesian and normalized polar coordinates respectively. The range of $\theta$ is $[0\ 2\pi]$ and $r$ is $[0\ 1]$. $x(r,\theta)$ and $y(r,\theta)$ are defined as linear combinations set of pupil boundary points. The formulas given in the following equations perform the transformation,

$$x(r,\theta) = (1-r)x_p(\theta) + x_i(\theta)$$
$$y(r,\theta) = (1-r)y_p(\theta) + y_i(\theta)$$
$$x_p(\theta) = x_{p0}(\theta) + r_p Cos(\theta)$$
$$y_p(\theta) = y_{p0}(\theta) + r_p Sin(\theta)$$
$$x_i(\theta) = x_{i0}(\theta) + r_i Cos(\theta)$$
$$y_i(\theta) = y_{i0}(\theta) + r_i Sin(\theta)$$

where $(x_p, y_p)$ and $(x_i, y_i)$ are the coordinates on the pupil and iris boundaries along the $\theta$ direction. $(x_{p0}, y_{p0}), (x_{i0}, y_{i0})$ are the coordinates of pupil and iris centers [39].

*3) Extraction of iris texture:* The normalized 2D form image is broken up into 1D signal, and these signals are used to convolve with 1D Gabor wavelets. The frequency response of a Log-Gabor filter is given as,

$$G(f) = \exp\left(\frac{-(\log(f/f_0))^2}{2(\log(\sigma/f_0))^2}\right)$$

Where $f_0$ represents the centre frequency, and $\sigma$ gives the bandwidth of the filter [41].

The Log-Gabor filter outputs the biometric feature (texture properties) of the iris.

*C. Fusion of Fingerprint and Iris Features*

We have at hand two sets of features namely, 1) Fingerprint features and 2) Iris features. The next step is to fuse the two sets of features at the feature level to obtain a multimodal biometric template that can perform biometric authentication.

*Feature Representation*: Fingerprint - Each minutiae point extracted from a fingerprint image is represented as $(x, y)$ coordinates. Here, we store those extracted minutiae points in two different vectors: Vector $F_1$ contains all the $x$ co-ordinate values and Vector $F_2$ contains all the $y$ co-ordinate values.

$$F_1 = [x_1\ x_2\ x_3\ \ldots x_n]\ ;\ |F_1| = n$$
$$F_2 = [y_1\ y_2\ y_3\ \ldots y_n]\ ;\ |F_2| = n$$

Iris - The texture properties obtained from the log-gabor filter are complex numbers $(a + ib)$. Similar to fingerprint representation, we also store the iris texture features in two different vectors: Vector $I_1$ contains the real part of the complex numbers and Vector $I_2$ contains the imaginary part of the complex numbers.

$$I_1 = [a_1\ a_2\ a_3\ \ldots a_m]\ ;\ |I_1| = m$$
$$I_2 = [b_1\ b_2\ b_3\ \ldots b_m]\ ;\ |I_2| = m$$

Thereby, the input to the fusion process (multimodal biometric generation) will be four vectors $F_1, F_2, I_1$ and $I_2$. The fusion process results with the multimodal biometric template. The steps involved in fusion of biometric feature vectors are as follows.

*1) Shuffling of individual feature vectors:* The first step in the fusion process is the shuffling of each of the individual feature vectors $F_1, F_2, I_1$ and $I_2$. The steps involved in the shuffling of vector $F_1$ are,

i. A random vector $R$ of size $F_1$ is generated. The random vector $R$ is controlled by the seed value.

ii. For shuffling the $i^{th}$ component of fingerprint feature vector $F_1$,

  a) The $i^{th}$ component of the random vector $R$ is multiplied with a large integer value.

  b) The product value obtained is modulo operated with the size of the fingerprint feature vector $F_1$.

  c) The resultant value is the index say '$j$' to be interchanged with. The components in the $i^{th}$ and $j^{th}$ indexes are interchanged.

iii. Step (ii) is repeated for every component of $F_1$. The shuffled vector $F_1$ is represented as $S_1$.

The above process is repeated for every other vectors $F_2, I_1$ and $I_2$ with $S_1$ $S_2$ and $S_3$ as random vectors respectively, where $S_2$ is shuffled $F_2$ and $S_3$ is shuffled $I_1$. The shuffling process results with four vectors $S_1, S_2, S_3$ and $S_4$.

*2) Concatenation of shuffled feature vectors:* The next step is to concatenate the shuffled vectors process $S_1, S_2, S_3$ and $S_4$. Here, we concatenate the shuffled fingerprints $S_1$ and $S_2$ with the shuffled iris features $S_3$ and $S_4$ respectively. The concatenation of the vectors $S_1$ and $S_3$ is carried out as follows:

i. A vector $M_1$ of size $|S_1|+|S_3|$ is created and its first $|S_3|$ values are filled with $S_3$.

ii. For every component $S_1$,

  a) The corresponding indexed component of $M_1$ say '$t$' is chosen.

  b) Logical right shift operation is carried in $M_1$ from index '$t$'.

  c) The component of $S_1$ is inserted into the emptied $t^{th}$ index of $M_1$.

The aforesaid process is carried out between shuffled vectors $S_2$ and $S_4$ to form vector $M_2$. Thereby, the concatenation process results with two vectors $M_1$ and $M_2$.

*3) Merging of the concatenated feature vectors:* The last step in generating the multimodal biometric template $B_T$ is the merging of two vectors $M_1$ and $M_2$. The steps involved in the merging process are as follows.

i. For every component of $M_1$ and $M_2$,

  a. The components $M_{11}$ and $M_{21}$ are converted into their binary form.

  b. Binary *NOR* operation is performed between the components $M_{11}$ and $M_{21}$.

  c. The resultant binary value is then converted back into decimal form.

ii. These decimal values are stored in the vector $B_T$, which serves multimodal biometric template.

*D. Generation of Cryptographic Key from Fused Features*

The final step of the proposed approach is the generation of the k-bit cryptographic key from multimodal biometric template $B_T$. The template vector $B_T$ can be represented as,

$$B_T = [b_{T_1}\ b_{T_2}\ b_{T_3}\ \ldots\ b_{T_h}]$$

The set of distinct components in the template vector $B_T$ are identified and are stored in another vector $U_{BT}$.

$$U_{BT} = [u_1\ u_2\ u_3\ \cdots\ u_d]\ ;\ |U_{BT}| \leq |B_T|$$

The vector $U_{BT}$ is then resized to $k$ components suitable for generating the *k*-bit key. The resize procedure employed in the proposed approach,

$$B = \begin{cases} [u_1\ u_2\ \cdots\ u_k] & ;if\ |U_{BT}|>k \\ [u_1\ u_2\ \cdots\ u_d] << u_i;\ d+1 \geq i \geq k & ;if\ |U_{BT}|<k \end{cases}$$

Where, $u_i = \dfrac{1}{d}\sum_{j=1}^{d} u_j$

Finally, the key $K_B$ is generated from the vector B,

$$K_B << B_i \bmod 2,\ i = 1,2,3\ldots k$$

IV. EXPERIMENTAL RESULTS

The experimental results of the proposed approach have been presented in this section. The proposed approach is implemented in Matlab (Matlab7.4). We have tested the proposed approach with different sets of fingerprint and iris images corresponding to individuals. The fingerprint images employed in the proposed approach have been collected from publicly available databases. The input fingerprint image, the extracted minutiae points and the intermediate results of the proposed approach are shown in figure 2. For iris feature extraction, we use iris images obtained from CASIA Iris Image Database collected by Institute of Automation, Chinese Academy of Science. The input iris image, the normalized iris image and the intermediate results of the proposed approach are portrayed in figure 3. Finally, the generated 256-bit cryptographic key obtained from the proposed approach is depicted in figure 4.

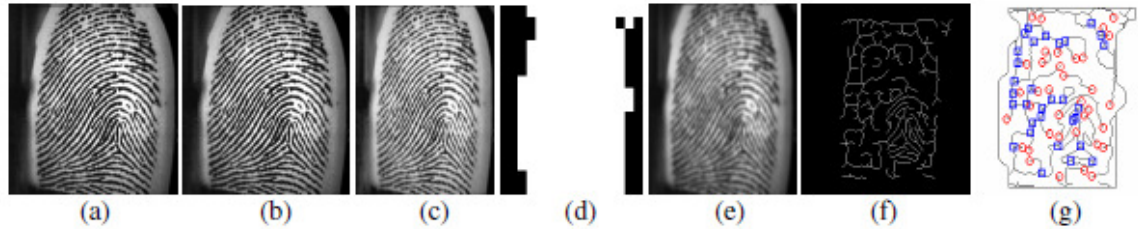

Figure 2. (a) Input fingerprint image (b) Histogram Equalized Image (c) Wiener Filtered Image (d) Segmented Image (e) Enhanced image (f) Morphological Processed Image (g) Fingerprint image with Minutiae points

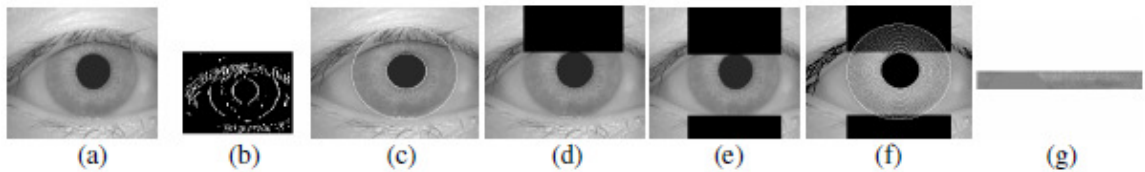

Figure 3. (a) Input Iris image (b) Edge detected image (c) Located pupil and iris boundary (d) Detected top eyelid region (e) Detected top and bottom eyelid region (f) Segmented Iris image (g) Normalized iris image

```
0101010101010001010101011010101001011100010010010101010101101010110010001000001000101000
100111011010100100000001000001000100001010101100000010100000010000100011000101010110010
0011010000100110000000000010100100010000000100000000010100000000100001000001000000010
```

Figure 4. Generated 256 bit key

## V. CONCLUSION

In this paper, we have attempted to generate a secure cryptographic key by incorporating multiple biometrics modalities of human being, so as to provide better security. An efficient approach for generation of secure cryptographic key based on multimodal biometrics (Iris and fingerprint) has been presented in this paper. The proposed approach has composed of three modules namely, 1) Feature extraction, 2) Multimodal biometric template generation and 3) Cryptographic key generation. Firstly, the features, minutiae points and texture properties have been extracted from the fingerprint and iris images respectively. Then, the extracted features have been combined together at the feature level to obtain the multi-biometric template. Lastly, a 256-bit secure cryptographic key has been generated from the multi-biometric template. For experimentation, we have employed the fingerprint images obtained from publicly available sources and the iris images from CASIA Iris Database. The experimental results have demonstrated the efficiency of the proposed approach to produce user-specific strong cryptographic keys.

**Authors Detail:**

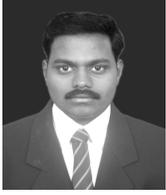

**Mr.A.Jagadeesan** was born in Coimbatore, India on June 14, 1979. He graduated from Bannari Amman Institute of Technology in 2000 with a degree in Electrical and Electronics Engineering. He completed his Master of Technology in Bio-medical Signal Processing and Instrumentation from SASTRA University in 2002. Thereafter he joined as a Lecturer in K.S.Rangasamy College of Technology till 2007. Now working as a Senior Lecturer in Bannari Amman Institute of Technology. He is a research scholar in the Department of Information and Communication Engineering. His area of interest includes Biometrics, Digital Image Processing, Embedded Systems and Computer Networks. He is a life member in ISTE and BMESI. He is also a member of Association of Computers, Electronics and Electrical Engineers (ACEE) and International Association of Engineers (IAENG).

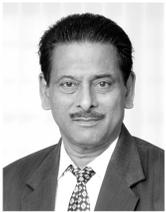

**Dr. K.Duraiswamy** received his B.E. degree in Electrical and Electronics Engineering from P.S.G. College of Technology, Coimbatore in 1965 and M.Sc. (Engg) from P.S.G. College of Technology, Coimbatore in 1968 and Ph.D. from Anna University in 1986. From 1965 to 1966 he was in Electricity Board. From 1968 to 1970 he was working in ACCET, Karaikudi. From 1970 to 1983, he was working in Government College of Engineering Salem. From 1983 to 1995, he was with Government College of Technology, Coimbatore as Professor. From 1995 to 2005 he was working as Principal at K.S.Rangasamy College of Technology, Tiruchengode and presently he is serving as Dean of KSRCT. He is interested in Digital Image Processing, Computer Architecture and Compiler Design. He received 7 years Long Service Gold Medal for NCC. He is a life member in ISTE, Senior member in IEEE and a member of CSI.